\documentclass[12pt]{article}
\usepackage[dvips]{graphicx}
\title{Physics of vacuum at ITEP and around}
\author{L.B.Okun \\
ITEP, Moscow, 117218, Russia \\ Email: okun@heron.itep.ru}
\date{}
\begin{document}

\maketitle

\begin{abstract}

Recollections about a few episodes from the history of physics of
vacuum, connected with the names of Pomeranchuk, Landau,
Zeldovich, Sakharov and Kirzhnits. The text of the talk will be
published in the Proceedings of the International Conference
``From the Smallest to the Largest Distances'', Tribute to Jean
Tran-Thanh-Van, May 24-26, 2001 (``Surveys in High Energy
Physics'', Taylor and Francis, 2002, v.16, No.3).

\end{abstract}

The people about whom I will speak -- Pomeranchuk, Landau,
Zeldovich, Sakharov, Kirzhnits  -- all of them have passed away. I
will also speak about their disciples. I will stress  the
importance of international exchange of ideas, the crucial role of
experiments on P and CP violation, the interconnection of particle
physics and cosmology.

These three themes are central for my friend Jean Tran-Thanh-Van
whose outstanding contribution to physics we are celebrating. They
are also central for ITEP with its unique atmosphere. The walls
around us have seen the people about whom I will speak and heard
their voices.

Lack of time does not allow me to touch upon other important
subjects connected with vacuum. They are partly reflected in the
text of my talk ``Spacetime and Vacuum as seen from Moscow'' which
has about 90 references (see ``2001: A Spacetime Odyssey'',
Proceedings of the Inaugural Conference of the Michigan Center for
Theoretical Physics, University of Michigan, Ann Arbor, 21-25 May
2001, M.J. Duff and J.T. Liu eds., World Scientific, Singapore,
2002; hep-ph/0112031).

\vspace{5mm}

\begin{center}

***

\end{center}

\vspace{3mm}

When I became a student of  Pomeranchuk in 1950 I heard from him a
kind of a joke that the Book of Physics has two volumes: v.1. is
``Pumps and Manometers'', v. 2. ``Quantum Field Theory''. Also  I
heard a statement that vacuum is filled with most profound
physical content, which for me was absolutely unclear at that time
and a little bit more clear now. In the middle of the 1950's
Pomeranchuk formulated his famous theorem about the equality of
cross-sections of particles and antiparticles on a given target.
In the framework of Regge theory this equality is determined by
exchange of a Regge pole with vacuum quantum numbers. This pole
was called later on pomeranchukon or pomeron.

In 1956 Pomeranchuk encouraged a group of young people at ITEP to
study the paper of T.D. Lee and C.N. Yang on parity
non-conservation and to think about it. As a result we (Boris
Ioffe, Alexei Rudik and myself) came to the conclusion that the
effects predicted by Lee and Yang would mean not only violation of
parity, but also violation of charge conjugation invariance
between particles and antiparticles. Somewhat later the same
result was obtained by Lee, Oehme and Yang.

\vspace{5mm}

\begin{center}

***

\end{center}

\vspace{3mm}

We discussed our paper with Lev Landau who usually visited ITEP on
Wednesdays, while we visited his Institute of Physical Problems on
Thursdays. This was a tradition. Landau was extremely skeptical
about the possibility of  parity violation.  That was during the
whole year  1956. The reason for him to be skeptical was that he
thought that if parity is violated that would have an imprint on
vacuum, because whilst vacuum influences particles, the particles
influence vacuum. Hence there should be some asymmetry in vacuum,
and the main idea of Landau was that vacuum should be completely
symmetric. Therefore he resisted all these attempts, though in
1950 he wrote a paper with Ginzburg on superconductivity in which
the notion of spontaneous symmetry breaking was brought into
condensed matter physics and a phenomenon, which was later
discovered in field theory and is called now Higgs mechanism, was
suggested. So, for him it would be quite natural to assume that
there is a spontaneous symmetry breaking in vacuum. But that's the
life, it is not always logical. Thus the fact is that Landau
resisted a vacuum which is not symmetric.

Suddenly, just overnight, he came with an idea that parity could
be violated, but it should be violated in such a way that
particles and antiparticles contribute with opposite sign to the
vacuum. And for that there should exist a CP-symmetry: C for
charge conjugation and P for parity mirror reflection. In that way
vacuum would be still symmetric and the whole sin of parity
violation will be with particles, not with vacuum. There will
exist left-handed neutrino and right-handed antineutrino, but
there would be no left-handed antineutrinos or  right-handed
neutrinos. That was his idea.

The idea that CP is an absolute symmetry was widely accepted, it
was beautiful. Personally I was extremely impressed by this
solution. However I couldn't understand what's wrong if I write a
Lagrangian and put in some places of the Lagrangian some
coefficients which are not real but complex, thus introducing
corresponding to CP-violation. Therefore I insisted that
experiments should be done looking for CP-violating processes. One
of them was proposed by Okubo with sigma and antisigma hyperons.

Another one was $K_L \to \pi^+ \pi^-$ an obvious candidate. The
search for this decay was undertaken in Dubna in the early 1960s
in a bubble chamber. They had about 600 3-body decays of $K_L^0$
and they looked for 2-body decays which was quite easy to
discriminate because of missing momentum in the 3-body decays.
It's either neutral pion in $\pi^+\pi^-\pi^0$ or neutrino in
semileptonic decays. And there would be just 2 charged pions going
back to back in the 2-pion decay. They collected 600 decays and
there was not a single 2-body decay. Then the Director of the Lab.
stopped this experiment as having no prospect, just waste of time.

If you look into the Particle data booklet, you will find that
each 350 3-body decays of $K_L$ are accompanied by one 2-body
decay of $K_L$. Thus by collecting 600 events if they were lucky
or if the Director were more patient, they would discover
CP-violation. But they were unlucky: the discovery came only two
years later and it was done by Fitch, Cronin, Turlay and
Christensen.

\vspace{5mm}

\begin{center}

***

\end{center}

\vspace{3mm}

Thus, the absolute CP-conservation  was disproved.

However the idea of Landau that vacuum must be symmetric pushed
Pomeranchuk,  Kobzarev and myself to consider a possibility
according to which to each of our particles corresponds another
particle with exactly the same properties -- same mass, spin,
charge -- everything. And the difference would be only in the
phase of CP-violation for these particles. So, what we postulated
is the existence of two terms of Lagrangian. One term is the known
particles, the other term is these mirror particles, mirror in the
CP sense.

As all interactions between the mirror particles are exactly the
same as between our particles, there should be mirror atoms,
mirror molecules, mirror condensed matter, mirror stars and even
mirror life. And all this would be unobservable, because there was
no interaction of the electromagnetic, weak, or strong type which
would connect these two worlds. The only interaction which
connected them (we were forced to this conclusion) was
gravitational one. Because if there were separate gravitons in the
two worlds, then it would be not a scientific assumption.  It's
just a matter of faith, but not of science. It couldn't be
scientifically proved or disproved.

There were many papers and they still appear now. But I am not
especially enthusiastic about it for many years, because it seems
to me there are other better ideas: namely, to accept the
asymmetry of vacuum and not to insist on the symmetry of it.

\vspace{5mm}

\begin{center}

***

\end{center}

\vspace{3mm}

My next subject is connected with cosmological term. All of you
know that cosmological term was introduced by Einstein in 1916 and
that later he considered it as a blunder and there were a lot of
discussions whether it should be or should not be. You certainly
heard  about them.

It is paradoxical, but it took many years before the first paper
appeared in which it was proven that it should be (not in a
mathematical way, but in physical, by Yakov Zeldovich). If you
have quantum mechanics, then there must be in the framework of
general relativity a cosmological term due to virtual particles.
In other words, due to zero state quantum oscillations.

Zeldovich was the first who published estimates of the
cosmological term. His first estimate was that it must be more
than 120 orders of magnitude larger than the upper limit on it.
Then he tried to invent various expressions which would contain
not $m_{Pl}^4$, but somehow to get $m_{Pl}$ in denominator to make
it proportional to the gravitational coupling constant.

Moreover  Zeldovich  was the first to notice and to publish in
1968 the observation  that the contribution of bosons and fermions
to the cosmological term are of opposite sign, and  therefore in
principle if there is boson-fermion symmetry, they would cancel
each other. If the symmetry is broken then they would cancel
partly.

This was a few years before the time when  Yuri Golfand and Evgeny
Likhtman suggested that there exists supersymmetry. In 1968 nobody
spoke about symmetry between fermions and bosons.

\vspace{5mm}

\begin{center}

***

\end{center}

\vspace{3mm}

Here I would like to make a digression and to tell you about
another piece of work done in an office which is adjacent to my
office at ITEP.

There in 1965 Mikhail Terentyev discovered together with  Vladimir
Vanyashin (Vanyashin is a physicist from Ukrain who still comes
from time to time to ITEP, while Terentyev died a few years ago at
the age of 60) that, unlike in quantum electrodynamics, in
nonabelian theories vacuum polarization has a negative sign. This
result has been derived again by Iosif Khriplovich in 1969. There
is so called ghost in the gauge theories. If you draw Feynman
diagrams, you cannot limit yourself with propagators of usual
particles only. You have to add propagators of ghosts which were
introduced by Fadeev and Popov in 1966. In 1965 they were not
known to Vanyashin and Terentyev. Therefore their result was not
22 in nominator, but 21.

This mistake was corrected by Khriplovich, who again not using the
ghost but using the so-called Coulomb gauge derived the correct
result. But neither they, nor the people around them realized what
a great discovery was made by these people. Personally I didn't
understand and my colleagues didn't understand, and nobody
understood at that time. The discovery became clear after
Gell-Mann introduced the notion of color, after QCD was formulated
and after Gross, Wilczek and Politzer  coined the term
``asymptotic freedom'': this negative sign in nonabelian gauge
theories leads to decrease of $\alpha$,  of the square of charge,
to zero at large momentum transfers. I have to admit that we were
not attentive enough to what was going on at that time in 1969 at
SLAC, to deep inelastic scattering, to partons.

\vspace{5mm}

\begin{center}

***

\end{center}

\vspace{3mm}

Another important contribution which was made at ITEP is the
so-called QCD sum rules, sometimes called ITEP sum rules. They
were written by Mikhail Shifman, Arkady Vainshtein and Valentin
Zakharov in  the late 1970s. They  exploited the fact that the
gauge coupling constant $\alpha$ not only decreases at short
distances, but also increases at large distances. And as a result
at large distances (large means $10^{-13}$ cm) the strong
interaction becomes really strong  and nonperturbative. One cannot
apply perturbation theory to it. They managed to write QCD sum
rules in which vacuum expectation values of gluonic fields and
quark fields were introduced. There are now, I believe, a few
thousand papers which develop this  theoretical discovery.

\vspace{5mm}

\begin{center}

***

\end{center}

\vspace{3mm}

Sakharov didn't work at ITEP, he was at Lebedev Institute. And
from Lebedev Institute he went to a secret city Arzamas-16 (now it
is called Sarov) where he became ``the father of the Russian
hydrogen bomb''. Then he returned back to Moscow in the 60s, just
after the discovery of CP-violation. He was practically uneducated
in particle physics, and it was astonishing to see how fast he
went to the heart of the subject.

I already mentioned that the decays of $\Sigma$ hyperons and
$\bar\Sigma$ hyperons are different  when CP is violated. Sakharov
used this difference as a prototype. What is the characteristic
for these decays? That strangeness is violated: $\Sigma$ hyperons
are strange while the products of their decay (protons and pions)
are nonstrange. Sakharov went from weak interactions with
strangeness violation directly to the idea of baryon number
violation and from $\Sigma$ decays to the Universe. He used a
similar mechanism for the early Universe in order to explain why
we are built from particles and not antiparticles. Or, better to
say, why there is not an equal amount of particles and
antiparticles in our world. In the latter case we would  not exist
at all, because particles and antiparticles would annihilate and
there would be nothing to build our matter from. Protons and
antiprotons, electrons and positrons would annihilate, and there
will be a world full of neutrinos and photons.

Thus, he started with hot symmetric Universe, big bang Universe.
And he explained how could it be that from this symmetric Universe
in which the number of particles and antiparticles were equal, our
world appeared. His idea was that this occurred due to
CP-violation, to baryon number nonconservation and to
non-equilibrium state in which this primordial soup was cooling
due to the expansion of the Universe. Because of these three
reasons there appeared a tiny excess of baryons compared to
antibaryons. This tiny excess survived because there were baryons
to which there was no match to annihilate with. While the majority
((majority means a billion of baryons and antibaryons) would
annihilate, one baryon would stay alone and survive. And that's
how he explained our existence.

I believe, this paper, only a few pages, is one of the most daring
papers of the last century.

The idea of baryon number nonconservation was necessary for
Sakharov, but it was not provided by the theory which existed at
that time. Later on a number of mechanisms were theoretically
discovered which lead to baryon number nonconservation. At present
there are several mechanisms, how to violate baryon number
conservation. But unfortunately they create new problems because
through them the asymmetry which could be produced at the early
stages of the Universe could be destroyed at the late stages.

\vspace{5mm}

\begin{center}

***

\end{center}

\vspace{3mm}

The name Sakharov is quite famous and the name Kirzhnits is not so
famous. David Kirzhnits was a colleague of Sakharov also at
Lebedev Institute. I remember how in 1972 he told me (we were at
the Conference in Tashkent) about his idea of phase transitions in
vacuum. That was absolutely new to me. I could understand face
transition in some material medium, but not in vacuum. At that
time the gauge symmetry $SU(3)\times SU(2)\times U(1)$ broken by
Higgs mechanism was already known. And the idea of  Kirzhnits was
that if we go backward in time, then at the stage of hot Universe,
at very high temperatures, the symmetry would be restored and all
massive particles would become massless. There would be no Higgs
vacuum expectation value in vacuum. This idea was further
developed later on by Kirzhnits and his student Andrei Linde.

A few years later Igor Kobzarev, Yakov Zeldovich and myself used
this idea of phase transitions when considering a model of
CP-violation proposed by T.D. Lee. In this model CP is
spontaneously broken. The Lagrangian is CP-invariant. However, its
potential has two minima: at one of them a neutral pseudoscalar
field $\phi$ has a classical solution $\phi = +\phi_0$ while
another one $-\phi_0$. This expectation values lead to
CP-violation with the sign of CP-violation being different in
these two minima. Thus, if the Universe chooses the first minimum,
when it cools down, the phase would be plus in all CP-violating
effects; if it chooses the second minimum, the phase would be
minus. But various parts of the Universe at early stages are not
casually connected, they cannot ``talk'' to each other. Therefore,
each point has to choose its sign for himself (or itself, herself
- I don't know). As a result you get a kind of a chessboard: you
have regions with plus sign and regions with minus sign. More
precisely it looks like a continent in which there are sees and
there are islands in the sees and there are lakes on the islands
and islands in the lakes and so on. All these different vacuum
domains are separated by walls. It's very easy to understand: if
you have in one point the first vacuum and in another point the
second one, you will get large potential energy between them. If
the border region is thick enough, you will have a lot of energy
in the border. If you try to squeeze this border, you will find
that you cannot squeeze it too much due to uncertainty principle.
You end up with an absolutely new kind of a solution in field
theory.

We are used to consider field theory in terms of particles,
microscopic objects. These borders, these walls, how we called
them, were even not macroscopic objects, they were megascopic
objects, they were huge. We considered the cosmology of Universe
filled with such walls and the cosmology turned out to be very
unusual: the expansion of the Universe was much faster than in the
Friedmann model: the wall which is the closest to us  had to go
far behind the horizon. But it could manifest itself by perturbing
the black body radiation. We don't seat in the center of our
domain, there is no reason for us to be at the center, we are
ordinary people. Therefore we are closer to one of the walls and
then there must be an asymmetry of black body radiation. But there
is no such symmetry in Nature.

The domain wall was the first megascopic solution in quantum field
theory. Soon afterwards other solutions appeared. A.Polyakov  and
G.'t Hooft  discovered monopoles in SU(2), T.W.B.Kibble discovered
strings, not microscopic strings, not superstrings, but
cosmological strings which go through the whole Universe. And
there is a whole theoretical industry of such solutions.

\vspace{5mm}

\begin{center}

***

\end{center}

\vspace{3mm}

In 1974 T.D.Lee and G.C.Wick considered another model in which the
potential of the field $\phi$ was slightly asymmetric. Therefore
one of the two minima was  lower and stable, while the other one
was higher and metastable. If the Universe were in higher place,
it could tunnel to lower place.

What does it mean: the Universe can tunnel? Assume that at a given
place you have the upper vacuum; then you will gain energy, if you
go at that place to the lower vacuum. Thus, at first sight a
simple question arises: Why not to go to a lower vacuum here and
now? The answer is that between two vacua there is a wall which
I've already described before: a material wall, very heavy, very
dense. You need energy to create it. Therefore there exists a
critical radius $R_c$ at which the gain in energy in volume of a
bubble is becoming larger than the loss of energy used for
creation of the surface of the bubble.

It was a very frightening experience which I had when I first
thought about these bubbles. I thought about the possibility that
at some collider the collision of two particles would enhance the
probability of creating such a microbubble. And if a bubble of
critical dimension is produced, then it can expand infinitely,
because the volume energy goes like $r^3$ and the surface energy
-- like $r^2$. Thus, for large $r$ volume will predominate. And
very soon the wall of the bubble would move with a velocity of
light, and the  bubble would expand and destroy the whole world. I
really shivered when I thought about this. But then I somehow
relaxed by thinking about the past: that Universe was hot, there
was a lot of various collisions in it. Therefore if bubbles could
be produced by collisions, then they already were produced and we
are living in a true vacuum now.

A few months later I told Andrei Sakharov about these bubbles. I
vividly remember his reaction. He said: ``Such theoretical work
should be forbidden. It's too dangerous''. I tried to persuade him
with arguments about past of the Universe and all that. And he
said: ``Nobody had collided two lead nuclei in the Universe''. I
took this quite seriously.

But in 1984 a paper by  P. Hut appeared, who considered the same
problem. Our conversation was in 1974 and 10 years later this
paper appeared. Hut had estimated a number of uranium-uranium
collisions in the Universe during all these billions of years of
its existence. Then other people estimated. They found (this was
done when RHIC at Brookhaven started to be discussed) that during
the existence of Universe there were enough  lead-lead, iron-iron,
uranium-uranium and even gold-gold collisions. There were about
$10^{47}$ iron-iron collisions, while RHIC would produce at most
$10^{12}$ through its lifetime. Thus, we are 30 orders of
magnitude safe.

I would like to conclude by saying that there was an enormous
progress during last 50 years in the understanding of physics in
general and physics of vacuum in particular. As a result of this
progress important problems have been solved, but the number of
unsolved problems facing us has increased immensely. I hope that
there will be even greater progress in the next 50 years.

Thank you.

\end{document}